\documentclass[11pt]{JHEP3}

\usepackage{epsfig}
\epsfclipon

\def\Dsl{\hbox{/\kern-.6700em\it D}} 
\def\dsl{\hbox{/\kern-.5300em$\partial$}}

\def\eq{\begin{equation}}
\def\eeq{\end{equation}}
\def\eqa{\begin{eqnarray}}
\def\eeqa{\end{eqnarray}}

\def\bd{\begin{displaymath}}
\def\ed{\end{diplaymath}}

\def\Box{ {\,\lower 0.9pt\vbox{\hrule\hbox{\vrule height0.2cm \hskip 0.2cm \vrule height 0.2cm }\hrule}\,}}

\def\lsim{{\ \lower-1.2pt\vbox{\hbox{\rlap{$<$}\lower5pt\vbox{\hbox{$\sim$}}}}\ }}

\def\gsim{{\ \lower-1.2pt\vbox{\hbox{\rlap{$>$}\lower5pt\vbox{\hbox{$\sim$}}}}\ }}

\def\g{\gamma}
\def\d{\delta}
\def\e{\eta}

\def\Dsl{\hbox{/\kern-.6700em\it D}} 
\def\dsl{\hbox{/\kern-.5300em$\partial$}}

\def\beginvector{\left( \begin{array}{c} }
\def\endvector{\end{array} \right)}
\def\endignore{}
\def\ignore#1\endignore{}

\def\S2{{\mathcal S}^2}

\def\de{\partial}
\def\a{\alpha}

\def\g{\gamma}

\def\d{\delta}
\def\D{\Delta}
\def\e{\eta}
\def\l{\lambda}
\def\L{\Lambda}

\def\m{\mu}
\def\n{\nu}
\def\r{\rho}

\def\s{\sigma}
\def\S{\Sigma}

\def\f{\phi}

\newcommand{\Ref}[1]{(\ref{#1})}

\newcommand{\be}{\begin{equation}}
\newcommand{\ee}{\end{equation}}
\newcommand{\bea}{\begin{eqnarray}}
\newcommand{\eea}{\end{eqnarray}}
\newcommand{\beqar}{\begin{eqnarray*}}
\newcommand{\eeqar}{\end{eqnarray*}}

\def\vev{{\it vev}}

\title{Selftuning and its footprints}

\author{Hans-Peter Nilles,\,\,Antonios Papazoglou\, and 
Gianmassimo Tasinato
 \\ Physikalisches Institut der Universit\"at Bonn \\
 Nussallee 12,  D-53115 Bonn, Germany
\\ 
\\E-mails:
 \email{nilles@th.physik.uni-bonn.de,\,\,antpap@th.physik.uni-bonn.de,\,\,
tasinato@th.physik.uni-bonn.de}}

\abstract{
We re-consider the self tuning idea in brane world models of finite volume.
We notice that in a
large class  of  self tuning models, the four dimensional physics is
 sensitive to  the vacuum energy  on the brane.
In particular the compactification volume changes each time the tension of the brane is modified: consequently,  observable constants, as the effective Planck mass or masses  of matter fields, change as well. We notice that  the self tuning mechanism and the stabilization mechanism of the size of the extra dimensions are generically in  apparent conflict. We focus on a self tuning model in six spacetime dimensions to  analyze how the above considerations are explicitely  realized.}

\begin{document}


\section{Introduction}

The cosmological constant problem
is  one of the most serious manifestations of the difficulty
to join together a quantum theory of fields
with a theory of gravitation. In the presence of a non-zero
cosmological
 constant,  it is not possible
to find solutions that preserve  Poincar\'e symmetry;
in particular,  even  
a relatively small contribution to the energy momentum tensor in the form 
of vacuum energy  curves the
  space-time significantly, in  deep conflict  with 
observations (for a review, see \cite{weinberg}).

 To reconcile  the particle physics models, whose
predictions for the amount of cosmological
constant are  of the order of $M_{Pl} \simeq 10^{18}$ GeV,
  with the quite small amount
 of the currently observed vacuum energy, of the order
of $10^{-3}$ eV~\cite{obs}, one has
consequently  to rely on an
 extreme fine tuning at each order in  perturbation theory.
In addition, the above 
fine tunings have to be repeated from scratch every 
time that a phase transition in the  cosmological
 history takes place.

\vskip 0.3 cm
\noindent
{\it Four dimensional approaches to the cosmological constant problem}
\vskip 0.3 cm

\noindent
Various approaches have been considered to attack the problem. The most
natural is maybe to look for a  {\it powerful unbroken 
symmetry},  
such as supersymmetry or conformal symmetry, that cancels the
 quantum contributions 
to the vacuum energy. However, the observed  world, at least up to the
electroweak scale,  does not respect
any of these exact symmetries: thus, they must be broken at
some scale bigger than a scale of the order of $100$ GeV. This is then
the scale one expects as natural for the cosmological constant:
it is clear that, in this picture, the problem
can be ameliorated, but it is still far from been solved.

As an alternative  to exact symmetries, in the eighties
some effort had been devoted in  building the so-called
{\it adjustment mechanisms} \cite{adjust}.  In these mechanisms, a scalar 
field is suitably coupled to gravity  and matter:
 its dynamics should respond to any contribution to the vacuum 
energy,  adjusting  its value to zero, corresponding
to the minimum of the scalar potential.
 There has not been, however, a  successful realization of
 this idea.   Weinberg~\cite{weinberg}, moreover, provided a no-go 
theorem that, under some general assumptions, 
states that the potential for the compensator field, which should adjust the 
vacuum energy to zero, has a runaway behavior. That is,
 there is no stationary point
 for the potential of the scalar
 that should  realize the adjustment, and thus the mechanism 
cannot work.

Other quite interesting possibilities rely on the observation that
 the (expected large) value of the cosmological constant
can be compensated by some {\it non-dynamical} field, like a 4-form 
$F_{4}$ in four dimensions.
 The constant value of
this field (or, in some approaches,
of a large number of these fields~\cite{BP}) changes via a process of 
membrane nucleation
\cite{abbott,BT}, up to a value that compensates the cosmological 
constant
to a degree sufficient to render
it compatible with observations. This value is generally fixed by anthropic 
considerations, based on Weinberg's arguments on galaxy formation
\cite{anthropic};
alternatively, the desired value of $F_4$, that  compensates
the cosmological constant, 
 can be  obtained
considering tunneling effects between metastable vacua,
each  corresponding 
to a different value of the 4-form~\cite{FMSW}.

\vskip 0.3 cm
\noindent
{\it Self tuning from higher dimensions}
\vskip 0.3 cm

\noindent
More recently, the possibility that our world is confined on 
a hypersurface (called {\it brane})
 embedded in a higher dimensional spacetime (called {\it bulk}), 
suggested to physicists
new possibilities to tackle the problem. 
Suppose, for example, that we live on a 
three-brane ({\it i.e.} a brane with three spatial dimensions) located
on a singular region of a  higher 
dimensional background. All the fields of the standard model (SM)
are, by some mechanism, localized on the brane, and only gravity
(besides other fields which are either heavy or couple
 sufficiently weakly to the SM ones) can propagate through the entire higher 
dimensional space.
 One could imagine a model in which
any contribution to the cosmological constant on the brane
(that can be regarded in this context as the  {\it tension}
of the same brane),
 is transmitted to {\it bulk parameters}, like integration constants,
in such a way that a four dimensional observer does not realize any change 
in the four dimensional geometry. 
In particular, suppose  that it is possible to find solutions
in  which the four dimensions, that correspond
to the observed ones, preserve   Poincar\'e symmetry,
regardless of the value of the brane tension.  Of course, a  
solution of the cosmological 
constant problem  is interesting
when  this transmission from brane to bulk
occurs {\it automatically}, without the necessity 
to re-tune bulk quantities by hand every time the vacuum energy
 in four dimensions changes. Models that 
 realize this fascinating idea are called {\it self tuning
 models}.

Let us note here that as we have defined the {\it self tuning} mechanism, we are interested on solutions that the Poincar\'e vacuum is preserved irrespectively of the brane tension. However, this does not exclude that there are {\it nearby curved solutions}. In other words, there can in principle exist solutions for neighboring values of some bulk parameters which result in a curved four dimensional space. A completely satisfactory solution to the cosmological constant problem should forbid such curved solutions to exist close to the Poincar\'e vacuum. Nevertheless, in the following we follow a less ambitious road and content ourselves with finding vacua which realize self tuning as defined above even if there exist nearby curved solutions.

A self tuning model can realize a compensation of the brane
tension {\it via} bulk quantities in various ways. In the discussion 
of the present paper, we would like
to separate self tuning models in two classes. 

 The first class contains models in which a change
in the brane tension is accompanied by a change of integration
constants relative to fields that live in the bulk (like, for example,
scalars or gauge fields), in such a way that the higher dimensional
geometry is {\it not} modified with respect to the original one.
To this class belong, for example, the
 earlier attempts to construct self tuning models 
\cite{silverstein,arkani}, although a careful analysis 
of these examples revealed that a fine tuning is actually required
to 
compensate  the brane tension in a consistent way \cite{forste}.

The second class is more general, and includes models in which 
a change in the brane tension can be  compensated by
integration constants  that govern the geometry of the bulk; 
in this case, it is possible to relate a change in the brane tension
to a modification of  the background geometry.
Examples that belong to this class include
recent attempts to construct self tuning models in six dimensions
 \cite{luty,carroll,navarro1,cliff,new}: in these models,
 a  three 
dimensional brane is located on a conical singularity of a six dimensional
space, and any change in the brane
 tension is  compensated by a change in the deficit angle 
of the cone, in such a way to maintain a Poincar\'e invariant
four dimensional subspace.

\vskip 0.3 cm
\noindent
{\it Consequences for four dimensional physics}
\vskip 0.3 cm

\noindent
The  question that we would like 
to address in the present paper is the following: does the
  compensation 
of the brane tension
{\it via} higher dimensional parameters
 leave some effect at the level of four dimensional
physics? In other words, is the low energy, four dimensional
observer somehow sensitive
to the mechanism of cancellation of the cosmological constant?

In general, when a compensation mechanism of a
change in the brane tension presents {\it fine tuning}
at the higher dimensional level, the same characteristic is 
expected to be inherited also at the four dimensional effective level,
when the procedure of dimensional reduction is properly 
done.

When, instead, a {\it self  tuning} mechanism can be realized 
in a higher dimensional system, we will show that, in general, well defined,
measurable quantities in the effective model,
 like the effective Planck mass, explicitly depend
on the brane tension: if the tension of the brane changes, 
 measurable quantities change accordingly.
This is due to the fact that when a change in the brane tension
is compensated by  a (self tuned) modification  in the brane geometry, 
at the same time the volume
of the extra dimensions 
changes its size. Since there exist four dimensional quantities, like  the
Planck mass, 
that are proportional to this volume, we can find, after determining
the dependence of the volume on the brane tension,
 the remaining effects of the self tuning mechanism in the 
four dimensional physics. In particular, we can determine in which cases 
the effects of the self tuning mechanism influence in a drastic
way four dimensional quantities, providing new phenomenological
constraints for these models.

\smallskip

Generalizing this observation,
we   point out 
 what we can call a  {\it conflict} between
 the self tuning mechanism and  a  
mechanism  that stabilizes the size of the extra dimensions. 
  Indeed, 
the condition to have  stabilized  extra dimensions contrasts with
the requirements at the basis of each self tuning example.

Consequently, of particular interest is the analysis of how 
 this conflict manifests itself at the four dimensional 
effective level, for models that automatically incorporate  
a stabilization mechanism. We will explicitly discuss this issue
for a six dimensional model, 
and provide  the effective dimensionally reduced
description of the self tuning mechanism. We
identify  the four dimensional 
parameters that compensate a change in the vacuum energy,
and  discuss how the four dimensional remnant  of  the
stabilization constraint,
that corresponds to the quantization
condition of a non-dynamical 4-form field,  influences the cancellation mechanism.

\vskip 0.3 cm
\noindent
{\it Organization of the paper}
\vskip 0.3 cm

\noindent
We start  with a general discussion on how self tuning models can be in
 principle realized in an extra dimensional setup.
Then, we  examine various examples of fine tuned 
and self tuning models, observing that the  avaliable self tuning models 
present a variation of the compactification volume as one changes the 
tension of the brane we are interested in. We focus on 
the six dimensional example and we discuss the consequences of the variation
 of the compactification volume, namely the variation of the gravitational 
constant, or the masses  of  matter fields, as a function of 
the brane tension.  We continue  discussing  under which conditions
contributions to the brane tension lead to drastic effects
at the four dimensional level. To conclude, 
we provide a four dimensional interpretation of the self
 tuning mechanism
in terms of a non-dynamical 
field, pointing out  the conflict between self tuning and the 
stabilization mechanism,  
as soon as the quantization condition of the relevant
 form is taken into account. At the end we present our conclusions and we
 comment on the prospects of the self tuning models in extra spacetime 
dimensions.

\section{Self tuning from higher dimensions}\label{selfhigher}

In this  section we will discuss, in a well determined 
framework, the idea of self tuning
from higher dimensions,  and how it can be realized.
Our aim  is to present 
a setup that is sufficiently general
to comprehend the various features that characterize self tuning
models. In  particular, our framework
 allows us to explicitly  discuss, at the end
of the section, the consequences
of self tuning mechanisms
 at the level of four dimensional
physics.

\subsection{The setup}

Consider a $D$-dimensional  background
 that corresponds to a solution of the equations
of motion relative to the following action
\be\label{Daction}
S_{D}=\int d^{D}x \sqrt{-g_{D}}\left[M_{f}^{D-2}\mathcal{R}_{D}
- \mathcal{L}(\f, g_{D}^{M N}) \right]\,,
\ee
where ${\cal L}$ is a  Lagrangian that describes the $D$-dimensional
fields that we denote
 with $\f$~\footnote{We  always work in the Einstein frame 
in the $D$-dimensional action.}. We now concentrate
on  solutions in which
the metric presents a  symmetrically warped form:
\be\label{Dmetric}
d s_{D}^{2}=e^{2 W(y)}~ g_{\mu \nu}(x)\, d x^{\mu} d x^{\nu}
+\g_{m n}
(y)d y^{m} d y^{n}\,,
\ee
where the $x$   coordinates span a four dimensional subspace and 
the $y$ coordinates a $d$ dimensional compact internal space ($D=d+4$).
 Additionally, we will assume that the bulk fields depend {\it only}
 on the $d$ extra dimensions $y$.
 We consider the metric of the $4$-dimensional space, $g_{\mu \nu}$, 
given as an initial and fixed
 input (it can be flat or curved space).

It is important to notice here  that in
general  the knowledge of the $D$-dimensional action is {\it not enough} 
to completely fix  {\it  integration constants} relative to
 bulk fields (for example $\f(y_0)$) or to metric components. 
This is generally a welcome feature for the self tuning models:
these  integration constants
 will be fixed only after specifying 
suitable  conditions at the boundary and on the singular regions of the space
time.

\smallskip

Let us consequently  add, to  the above  $D$-dimensional
action,  {\it localized terms}, that describe
 three dimensional branes (labeled by an index $i=0,1, \dots n$),
placed 
in {\it singular surfaces} of  the bulk 
background geometry~\footnote{One 
can generalize our subsequent arguments for the case when
 branes of different dimensionality than three are also present.}.
 Their action is given by (notice that we use the {\it projected} metric in
the four dimensional space):
\be\label{4dact}
S_{4} = -\sum_{i=0}^n e^{4 W(y_{i})}  
 \, e^{\sigma_{i} \phi(y_{i})}\,T_{i} \int
d^{4}x \sqrt{-g(x)} \,.
\ee
For our arguments, we will only consider branes that 
contain tension, labeled by $T_{i}$. We also allow a direct coupling 
between bulk scalars and brane tensions, and we model this coupling 
 with  exponentials. The inclusion of these localized
 terms in general  fixes
completely  the integration constants of the general bulk 
solution.

\subsection{Self tuning at work}\label{sftuneaction}

\noindent
In this setup, the idea of the  self tuning
approach is the following.
Let us suppose  we live on the  brane situated
at $y=y_0$, with tension $T_{0}$.
 Then, it is  possible
to find solutions of the equations of motion in the bulk, that
respect the conditions given by  boundary terms
and maintain the same metric $g_{\mu \nu}$  for the four dimensional
slice,
in such a way that  there is {\it no fine tuning}  between $T_{0}$,
 the
tension of any other brane $T_{i}$, and/or  higher dimensional parameters
explicitly  appearing in $\mathcal{L}$. 

This means that
we can freely change the tension of our brane $T_{0}$, with the only 
consequence  that integration constants
relative to  bulk fields
 or to metric components {\it change accordingly }
 in such a way that the observed geometry
 in
four dimensions, described by $g_{\mu \nu}$,   is not modified. The 
simplest case corresponds to  the Minkowski metric $\e_{\m\n}$:
this is the case in which we will concentrate in the following, 
unless otherwise
stated.

\smallskip

Since we are looking for solutions with vanishing vacuum energy, the numerical
 value of the total action, evaluated on the solution, 
 should be zero:
 indeed, integrating out the extra dimensions,
 it is easy to see that this number corresponds
 to the
 value of the vacuum energy from 
 the  four dimensional effective point of view.
 Let us rewrite the total action 
in a way that facilitate our following discussion:
\bea\label{totact}
0 &=& S_{tot} = S_{D}+S_{4} \nonumber \\
&=&  S^{reg}_{D}+ M_f^{D-2}\sum_{i=0}^n
e^{4W(y_i)}R^{sing}_i \int d^{4}x \sqrt{-g}
-  \sum_{i=0}^n e^{4 W(y_i)}  
 \, e^{\s_i \phi(y_i)}\,T_{i} \int
d^{4}x \sqrt{-g}~~~~~
\eea
In the last equality of the previous
formula, the bulk action has been split into a regular part
that describes the higher dimensional curvature scalar and  the bulk 
fields, and a singular part on the scalar curvature, which  is present 
since  the 
branes are located at singular points of the geometry. For simplicity,
we will limit the following analysis to cases in  which 
up to two branes are present in the localized part of the action.
 We have written the singular part
 of the curvature scalar as
 ${\cal R}^{sing}_i=R^{sing}_i~\D^{(D-4)}(y-y_i)$ 
where $\D^{(D-4)}(y-y_i)$ is the generalized delta 
function in the curved space \cite{myers}~\footnote{This function
satisfies the following identity: $\int \sqrt{\g}\D^{(D-4)}(y-y_i)=1$.}.

As we have noted, the numerical value of the total action 
$S_{tot}$ cannot vary  keeping the 
Minkowski vacuum as solution in four dimensions: 
this should remain true also after modifying
 $T_{0}$. 
Thus, there ought to be cancellations between the above
  terms in  the action when $T_{0}$ changes. We will divide our discussion 
into two cases, regarding  how these cancellations occur. 
The first case corresponds to the special situation in which
 the  bulk part $S_{D}$ (that includes also the singular part of
the curvature) and the brane part $S_{4}$  remain separately invariant
after a change in $T_0$.
The second case describes the more general situation in which
 each of them can vary, but in such a way that their sum cancels to zero.

\subsubsection{Inter-brane compensation}\label{separate}

Let us start considering the possibility that both $S_{D}$
 and $S_{4}$, in (\ref{totact}), are {\it separately invariant} when 
the brane tension is modified. 
In such a case, a change in $T_{0}$ (the tension of the brane in which the SM
 fields  live)  should be 
entirely compensated by {\it brane} quantities,
that is, quantities that appear in $S_4$.  In particular,
we ask that the integration constants
 that compensate a change in the brane tension  do not
modify the bulk action $S_D$, and consequently do not change the
bulk geometry.

The natural quantity that
 acts as compensator of a change in $T_{0}$ is the value of
 $\phi(y_0)$, that
 is generally determined up to some integration constant.
The integration constants relative to the scalar, in this case,
change in such a way that they compensate a change in $T_0$, without 
changing the value of $S_D$. However, the scenario is potentially problematic, since in general
 a change on the integration constants of the bulk fields require,
for the consistency of the system, a corresponding change  
in the tension $T_1$ in a fine tuned way.

\smallskip

This is precisely what happens in the first attempts to realize this idea, in the five dimensional dilatonic models of \cite{silverstein,arkani}. As we will show in detail in  the next section
(also more complex frameworks  share the same problem), a change in
the integration constants, that allows to compensate a change in the
brane tension $T_{0}$,  requires also  a change on the tension of the
second brane $T_1$ \cite{forste,hebecker}.  We are not aware
of a complete, consistent model, belonging to this class,
that presents a self tuning behavior.

\subsubsection{Brane-bulk compensation}\label{mixed}

An alternative way to realize self tuning is the following.  Imagine
that any variation of the brane tension $T_0$  is accompanied by a
variation  of the bulk action  $S_D$ in (\ref{totact}), in such a way
that the
resulting value of the total   action remains null. If this
 condition
can be satisfied without fine tuning, and preserve a flat four
dimensional slice, one obtains a self tuning
mechanism to cancel the
cosmological constant {\it via} a change in  the geometry 
of the internal dimension.

\bigskip
We can have the case in which both the  {\it regular} and the {\it singular}
 parts of the bulk
action change with the brane tension.
This is actually what happens in the Randall-Sundrum models \cite{RS},
that however compensate the cosmological constant at the price of fine
tuning bulk parameters. Another example  
is the model of \cite{HM},
 where a 3-form gauge field is considered,
 with Lagrangian of the specific form $1/H^2$,
where $H$ denotes the 4-form 
field strength of that field. In this case, the brane
contribution is canceled by a correlated change of the bulk geometry
and of an integration constant of the bulk gauge field  solution,
 and no fine tuning of parameters is apparently required.

\bigskip 
In another situation, the regular part of the  bulk action remains
unchanged after a change on $T_{0}$, and 
the {\it singular part  of the bulk action alone} 
 compensates a change
in  the brane tension. As a consequence, only the
global properties of the geometry result modified, 
while locally the geometry remains the same.
Examples that belong to this class are  constructed in a six
dimensional background, that is the only dimensionality in which the
idea can be applied.
Indeed, Einstein 
equations (see Appendix \ref{appSing} for details) imply the following
equality, when the  regular part of the Einstein tensor cancels exactly 
the bulk part of the energy momentum tensor:

\be
M_f^{D-2}R^{sing}_i={4 \over D-2}e^{\s_i \f(y_i)}T_i \,.
\ee
Then, for the cancellation to be realized, one should have~\cite{luty}
\be
{4 \over D-2}=1 ~~~ \Rightarrow ~~~ D=6\,.
\ee
[In general if the $x$-coordinates span a $(p+1)$-dimensional space, 
the requirement would have been that $D=(p+1)+2$.]

Since it is possible to analyze  quite generally
how the compensation applies in this specific 
six dimensional situation, let us 
end this subsection with a discussion of how self tuning
is in principle  realized in this case, 
showing explicitly how the global properties
of the background change.

Since the mechanism of cancellation relies only on
 bulk quantities, we can, for the sake of simplicity, neglect  
the presence of other singular objects apart from our
 brane~\footnote{We do not mean that the presence of other
branes is not important for the consistency of the model. Simply, they are not
essential 
  for the features of the self tuning mechanism
 we intend to discuss.}.
In the simplest situation, let
 us impose that the six dimensional space time 
presents
an {\it azimuthal symmetry} around an axis. An ansatz for the metric
that preserves the four dimensional Poincar\'e symmetry is

\be\label{sixdmetric}
d s^{2}_6=e^{2W(r)} \eta_{\mu \nu} d x^{\mu} d x^{\nu}
+d r^{2} + \rho^{2}(r) d \f^{2}\,,
\ee
where,
in the previous formula, $r$ is a radial variable, while $\f$ an
angular variable with values $\f \in [0,2\pi)$.

Assuming that $\rho(r) \propto r$ for $r \to 0$, 
the point $r=0$ corresponds to a single
 point in the transverse space.
 A three-brane of tension
$T_{0}$, that extends
along the directions $x^{\mu}$ and is located in $r=0$, 
induces a conical singularity with deficit angle $\d$, related to the
brane tension  $via$ the equality
\be
{\d}  = \frac{T_{0}}{2M_{f}^{4}}\,.
\label{defang}
\ee

At this point, consider a change in the brane tension $T_0$.  A new solution
that takes into account this modification, and that preserves the required
azimuthal symmetry, differs from the previous one just by a change in
the value of  the deficit angle 
$\delta$,  that adjusts to the new value of the tension.
A change on this deficit  angle, indeed,  represents
a  modification of the global geometry.
If this change  does not imply a fine tuning on the
parameters appearing in the action, a self tuning mechanism is realized.

\subsection{What happens
to four dimensional physics?}\label{pause}
After our general discussion of how self tuning  can 
in principle be realized,
let us ask the following question: does the higher dimensional mechanism
to cancel the cosmological constant  have some effects
at the level of four dimensional physics?

In general, one expects that some remnant of the higher dimensional
mechanism should be present in the effective four dimensional
description.
Indeed, conceptually, the cosmological constant problem is present also 
at low scales, even lower than the typical energy scales in which higher
 dimensional physics manifests itself. Consequently, for this range 
of energies, 
one expects to find a 
four dimensional effective description for  the cancellation mechanism,
 although maybe quite non standard.

\smallskip
When the cosmological constant is compensated, in the higher dimensional 
framework, by some symmetry, one expects that some remnant 
of this symmetry is left at the four dimensional level, and the mechanism
of cancellation can  be effectively understood in terms of this inherited 
symmetry.
Instead,  when the responsible of the cancellation is a dynamical field
in the bulk,  one expects  some very  light degree of freedom 
to be present at the effective four dimensional level, that
dynamically compensates the cosmological constant.
Alternatively, the higher dimensional cancellation mechanism can
be ultimately 
 due to some non-dynamical parameter in the bulk, whose  value 
compensates the brane tension. In this case, one expects  that, 
at the four dimensional level,  the cancellation mechanism can be described 
in terms of some non-dynamical field. 

\smallskip
As we have seen, in general the self tuning mechanisms that
we have outlined are based on the last possibility:
the brane
tension is compensated by non-dynamical parameters in the bulk, like
integration constants. Naturally, the determination
of the effective four dimensional description
of the mechanism  is not an easy task.
For this reason, in the following we concentrate  our discussion  
 on the   effects of the higher dimensional mechanism
 at the level of the four dimensional
effective theory. Once these remaining effects
are identified, the connection between higher and
lower dimensional descriptions should become  clearer.

\smallskip

In general,  one expects 
that  models that require a fine tuning at the level of higher
dimensional parameters, present a fine tuning in the four dimensional 
effective action, after the procedure of dimensional reduction
is performed: the four dimensional model presents, at the effective 
level, the cosmological constant problem in its traditional 
form.

\smallskip

On the other hand, when we look at four dimensional effects of
 the most promising   self
tuning models,  belonging  to the class discussed
in
 Section
(\ref{mixed}), we know that they
 require a modification 
of the bulk geometry.
 We observe that  in this case  a dimensionful 
parameter of the four dimensional action varies when the brane tension 
changes: this is the effective Planck mass in four dimensions.
This is due to the fact that, in the examples
we discuss, a change of the geometry implies
 a modification in the volume of the internal dimensions.
In general, the volume is proportional to the effective Planck mass in four 
dimensions, and this quantity changes accordingly with the brane tension: as
a consequence,  an analysis of 
these models
should take  into account also  phenomenological 
constraints related to variations of the Planck mass.

In more general terms,
 in order for an extra dimensional  model to satisfy
 phenomenological constraints
on the behavior of gravitational interactions, it
 should incorporate a  mechanism
 that stabilizes  the size of the 
extra dimensions (at least at the present cosmological 
epoch).
 The self tuning models we consider require
just the opposite, since 
the volume
of the extra dimensions must  change in order  to compensate
a change in the brane tension.
As we will see  analyzing explicit examples, the requirement to maintain a
 fixed volume  for the extra dimensions imposes an additional constraint that
ruins the self tuning mechanism. This means that
whenever the higher dimensional
system includes  a stabilization mechanism,
there is a conflict  between self tuning and 
stabilization. This  will be particularly 
transparent in a six dimensional example, on which  we will
focus  our attention in Section (\ref{consequences}), presenting also a four dimensional description of this fact.

\section{Explicit examples}\label{secexamples}

In this section, we present  examples of 
higher dimensional
models, already known in the literature,
 in which the general arguments we have discussed so far find
an explicit  realization.

We start with a model in which the cancellation occurs between
the brane part of the action, that is, a model that belongs to the class discussed
in 
Section (\ref{separate}). We continue with three models that realize
a compensation using  quantities related to the bulk
geometry, and that consequently belong to
the class of Section (\ref{mixed}).

Finally, somehow outside the discussion of the previous section,
 we present a five dimensional self tuning model whose  bulk metric 
does not satisfy the ansatz (\ref{Dmetric}) but is instead
{\it asymmetrically 
warped}, in a sense we will
define below.  Also in this case the
 compensation occurs due to bulk quantities relative to the higher
 dimensional geometry, and the bulk volume changes.

\subsection{A five dimensional dilatonic model}

In this subsection, we discuss one of the dilatonic models 
of \cite{silverstein,arkani}. In this approach, a cancellation
mechanism is realized 
exploiting the properties of an integration constant 
relative to a bulk scalar field, without involving
quantities relative to the bulk geometry. The model 
belongs to the class discussed in Subsection (\ref{separate}),
and we will see how the arguments developed in that context
apply here.

 The five dimensional
 background
consists of a four dimensional Minkowski space
plus a compact extra dimension, modded by a ${\mathbb Z}_2$ symmetry
with two fixed points (located at the positions
$0$ and $y_s$) on which two three-branes are located.
We parameterize the five dimensional metric in the following way:

\be\label{ansfive}
d s_{5}^{2}=e^{2 W(y)}~ \eta_{\mu \nu}\, d x^{\mu} d x^{\nu}
+d y^2\,.
\ee

The model contains gravity, a bulk 
scalar field, and two branes, characterized by their tension and 
conformally coupled to the bulk scalar. The action
is given by:
 
\be\label{silveraction}
S_{tot}=M_f^3\int_{-y_s}^{y_s} dy d^4x \sqrt{-g_5}[{\cal R}_5
-{4\over 3}(\de \f)^2]-T_0\,e^{-{4 \over 3}\f(0)}\int \sqrt{-g}
-T_1\,e^{-{4 \over 3}\f(y_s)}e^{4W(y_s)}\int \sqrt{-g}
\ee

\noindent
where we have normalized $W(0)=0$.
The solution that maintains a four dimensional
 Poincar\'e vacuum, results:
\be
W={1 \over 4}\ln \left|{T_0 \over 3 M_f^3} e^{-{4 \over 3}\f_0} |y|-1\right|
\ee
and
\be
\f={3 \over 4}\ln \left|{T_0 \over 3 M_f^3} 
e^{-{4 \over 3}\f_0} |y|-1\right|+\f_0\,.
\ee
Notice that the solution for the scalar field is defined up to
an integration constant $\phi_0$: this observation
is crucial for the cancellation mechanism we are going to discuss.
The solution has a singularity in
 $y=y_s={3M_f^3 \over T_0}e^{{4 \over 3}\f_0}$, and 
 the r\^ole of the second brane in $y_s$ is just to screen
 the singularity \cite{forste}
 (for a general discussion about the presence
of background singularities in self tuning models
see also \cite{cline}). The 
boundary conditions on the second brane force to choose $T_1=-T_0$.

Suppose, now, that we live on the brane at the origin, and we vary
the tension $T_0$. To have a solution that, after dimensional
reduction, gives us a flat four dimensional geometry,
we must ask that the value of $S_{tot}$ does not change varying $T_0$,
and in particular it keeps a vanishing value.

Looking at the second term in the action (\ref{silveraction}), 
we see that any change in $T_0$ like  $T_0 \to e^{\l}T_0$
can be compensated by a change in the integration
constant 
$\f_0 \to \f_0+{3 \over 4}\l$. A change in $\phi_0$ does not
affect the bulk part of the action, that is,
 the first term in (\ref{silveraction}).
However, it is clear that the value of the third term changes
after this procedure, unless we also impose
by hand the additional condition $T_1 \to e^{\l}T_1$.
Thus, 
 we must  re-fine tune a parameter that appears explicitly in the action
to  maintain a   Poincar\'e invariant four dimensional subspace,
and the model presents an apparent fine tuning at the five dimensional 
level.

\smallskip

\subsection{The Randall-Sundrum model} 

In the (one brane) Randall-Sundrum model \cite{RS},  a three-brane
is located 
at the fixed point (located at the origin)
 of a ${\mathbb Z}_2$ symmetric $AdS_5$ background.  
The action for the model, that describes gravity, bulk
cosmological
constant, and brane tension, is given by

\be\label{actrand}
S_{tot}=\int_{-\infty}^{+\infty} dy d^4x
 \sqrt{-g_5}(M_f^3{\cal R}_5-\L_{B})-T_0\int d^4x \sqrt{-g}\,.
\ee

Considering formula (\ref{ansfive}) as ansatz for the metric, a 
solution that preserves four dimensional Poincar\'e 
invariance for any choice of the brane tension
$T_0$ is given by 
 $W(y)=-ky$, with  the parameter $k$ given by
 $k =\sqrt{-\L_B \over 12 M_f^3}$.
It is well known, however, that this mechanism of cancellation requires 
a fine tuning between brane tension and bulk cosmological constant. 
Indeed, the following condition must be imposed between brane and
bulk cosmological constant
\be
T_0=\sqrt{-12\,\Lambda_B \, M_f^3}.
\ee

 Any change of $T_0$ must be accompanied,
to maintain a four dimensional Poincar\'e space, by a change
of the five dimensional bulk cosmological constant, changing the 
value of the regular part of the bulk action in (\ref{actrand}).
Consequently,  this  model, that gives a (fine tuned) compensation
of the cosmological constant,
 belongs to the class discussed
in Section (\ref{mixed}). 

\smallskip
Notice also that this is the first model in which the higher
dimensional  mechanism to cancel the cosmological constant,
although it requires fine tuning, presents a clear footprint 
at the level of four dimensional physics, since it requires
 a modification of the bulk volume. Indeed, in this model,
the four dimensional Planck mass is given by
\be
M_{Pl}^{2} \,=\, \frac{M_{f}^{3}}{k} 
\,=\, \frac{12 M_{f}^{6}}{T_{0}}\,,
\ee
and a change in $T_{0}$ implies a change in the Planck mass.

\subsection{A six dimensional  model}\label{asixd}

We would like to discuss now the six dimensional model
of  \cite{carroll} (see also
\cite{navarro1,cliff,navarro2}). In this model, 
a three-brane is located on the singular point (a conical singularity) 
of a six dimensional background containing gravity,
a 2-form field strength, and cosmological constant~\footnote{The
system can be generalized including also  scalar fields: see
 \cite{gibbons,new}.}.
The action one considers is 

\be\label{sixdtotac}
S_{tot}=\int d^6x \sqrt{-g_6}\left[M_f^4{\cal R}_6
-\L_B -{1\over 4}F_{MN}^2\right]-T_0\int d^4x\sqrt{-g}\,.
\ee

\noindent
Looking for un-warped solutions
that maintain  a four dimensional
subspace with Poincar\'e symmetry, one finds as a solution  for the metric 
with a  locally spherical 
geometry in the two extra dimensions
\be\label{sixdfmetr}
d s_{6}^{2}=\e_{\m\n}dx^{\m}dx^{\n}
+R_{0}^{2}\left( d \theta^{2}+
\alpha^{2} \sin^{2}\theta \, d \phi^{2} \right)\,,
\ee

\noindent
while a solution for the 2-form is given by the following expression
\be\label{sixdgaf}
F_{\theta \f}=f\,\epsilon_{\theta \f}\,.
\ee
In the last formula, $f$ is a constant, while 
$\epsilon_{\theta \f}$ is the volume form of the two extra dimensions.
In the metric solution (\ref{sixdfmetr}), to obtain a flat
four dimensional background one must impose

\be\label{dkappa}
{ 1 \over R_0^2}={f^2 \over 2M_f^4}\hskip 1 cm , \hskip 1 cm \Lambda_{B}=
\frac{f^2}{2}\,\,.
\ee

 The important observation,
in this context, is that the formulas (\ref{dkappa}) do not depend on the brane
tension. The only dependence on $T_0$, in formulas
(\ref{sixdfmetr})-(\ref{dkappa}), is limited to the size of the deficit
angle, which is  related to $\alpha$. 
Indeed, one finds that
the sphere presents  conical singularities at the poles~\footnote{
One of the two conical singularities can be discarded by imposing
a ${\mathbb Z}_2$ symmetry  at the equator of the sphere.}, with deficit
angle $\delta$ given by 
\be\label{fordef}
\delta \equiv 2\pi(1-\alpha)=\frac{T_{0}}{2M_{f}^{4}}\,.
\ee

As a consequence, changing the tension $T_0$, one finds a new solution
simply changing  accordingly the deficit angle, following formula 
(\ref{fordef}).
Only the singular part of the curvature  changes, and the model belongs
to the class discussed in Section (\ref{mixed}).
In this way, the curvature of the four dimensional subspace remains invariant,
and, apparently, no fine tuning is required on quantities that
explicitly appear in the bulk action. 

\smallskip

At the four dimensional level, the self tuning mechanism 
has an important effect. Indeed, as we mentioned
in  
 Section (\ref{pause}), a change in the deficit angle implies
a change in the bulk volume. Consequently, the 
effective,
four dimensional Planck mass changes:
this observable quantity  results to 
 depend on the brane
tension as

\be\label{varsymm}
M_{Pl}^2(T_0)=4\pi\left[1-{T_0 \over 4 \pi M_f^4}\right]
R_{0}^2 M_{f}^{4} \equiv \a M_{Pl}^2(0)\,.
\ee
where in the last equality we defined, 
using also (\ref{fordef}),
the effective Planck mass in absence of branes on the singular points.
From the previous expression, it is apparent
that if one imposes to the
system the additional condition to maintain the same Planck mass after 
changing the brane tension (that is, keeping the volume fixed),
 one must also change the radius
$R_0$ of the sphere, and consequently also
 the bulk cosmological constant $\Lambda_B$ (see formula
(\ref{dkappa})).
In other terms, one ends with a fine tuning {\it \`a la} Randall-Sundrum.

Another important observation
concerning this model, that as we will see is related to  the 
issue of the variation of the Planck mass,
 has been made in 
 \cite{navarro2} (see also \cite{cliff}).
 There, it has been noticed that, for consistency
of the model, 
the monopole charge must satisfy 
a quantization condition that ruins the self tuning property:
the brane tension can acquire  only well
defined, quantized values, and cannot vary in a continuous way.

Both these two observations, and an interesting relation between them,
 will be re-considered and expanded
later on, in Section (\ref{consequences}).

\subsection{A five dimensional 3-form model}

The next
  model we will discuss, that 
has a self tuning behavior, is the one presented
in \cite{HM}.  It uses a 3-form field in five
 dimensions that are compactified on an ${\mathbb R}/{\mathbb Z}_2$ orbifold.
 The action for the 3-form field has the  unconventional form 
of $1/H^2$ where $H_{MNKL}$ is the field strength of the 3-form field. 
Since a 4-form field (here $H$) in five dimensions is dual to a
 scalar field $\s$, we can work
with  the equivalent action for that dual scalar field (as
done in the second reference of  \cite{HM}):
\be \label{HMact}
S_{tot}=\int_{-\infty}^{\infty} 
dy d^4x \sqrt{-g_5}\left[M_f^3{\cal R}_5
-\left(\de_M \s \de^M \s\right)^{1/3}-\L\right]-T_0\int d^4x \sqrt{-g}\,.
\ee

The solution that presents  a four dimensional
 Poincar\'e vacuum is the following, in terms of the ansatz
of formula (\ref{ansfive}):
\be
W(y)={1 \over 4}\ln \left[{\cosh c \over \cosh (4k|y|+c)}\right]\,,
\ee
while a solution for the scalar is 
\be
\s'(y)={216~ k^3 M_f^{9/2} \over  \cosh^3 (4k|y|+c)}\,,
\ee
with $k=\sqrt{-{\L \over 12 M_f^3}}$,
 $c=\tanh^{-1} \left(T_0 \over 12 k M_f^3\right)$, and where
 we have normalized to $W(0)=0$. This model belongs to the class
 discussed in (\ref{mixed}), and a variation 
of the brane tension $T_0$
 is compensated by a change in the integration constant
$c$, without, at least apparently, involving a fine tuning of other parameters.

Also this model, in any case, realizes self tuning at the price
of changing the volume of the extra dimensions, and consequently 
of the Planck mass. This quantity is given by  the following
formula, and depends on $T_0$ since it depends on $c$:
\bea
M_{Pl}^2(T_0) &=& 2 M_f^3  
\sqrt{\cosh c} \int_0^{\infty}{dy \over \sqrt{\cosh(4 k y+c)}}\, \nonumber\\
 &=& 2 M_f^3 \int_0^{\infty}{dy \over \sqrt{\cosh{4 k y}+
\sinh{4 k y}\tanh{c}}}
\eea

\smallskip
Let us end this subsection noticing
 that the present model is quite unconventional, 
 since, for example, the 
 term of the action that contains the derivatives
of the scalar contains a fractional  power of a quantity that
is not necessarily positive definite.

\subsection{An asymmetrically warped model}

To conclude our presentation
of higher dimensional examples to cancel the cosmological 
constant, we present another case 
 of self tuning model in five dimensions.  We consider 
a bulk geometry of the form~\cite{kraus,bowcock}:
\be
d s_{5}^{2}=-h(r) d t^{2}+\frac{d^{2} r}{h(r)}+r^{2} d\Omega_{3,k}^{2}\,.
\ee
where $d\Omega_{3,k}$ denotes a maximally symmetric three 
spatial surface of constant
curvature $k$, where the SM fields are supposed to live.
This form for  the geometry in five dimensions naturally appears
when we consider  backgrounds
that contain horizons  covering  naked singularities, a situation
that we consider here in order to avoid the introduction
of a second brane.
The model is called asymmetrically warped since time and the  three
spatial
dimensions are  differently  warped.

Since  this form for the metric is different from the one
discussed in the previous section, this example
 requires a separate discussion. In any case,
we will see that, also in this context,
 self tuning is obtained at the price
of changing the Planck mass when the brane tension changes.

\smallskip

In this background, a  brane-world is defined in the following
way. Let us  consider a hypersurface that resides on a fixed point
of a ${\mathbb Z}_{2}$ symmetry, and that extends 
along the three spatial dimensions
$\Omega_{3,k}$.
 Its  position on the other coordinates
$t$ and $r$ is   given in terms of  the proper parameter $\tau$:
\be
X^{\mu}=(t(\tau), r(\tau), x_{1}, x_{2}, x_{3})\,.
\ee
Defining in a suitable way the normal to the brane, 
and writing the  projected metric in the hypersurface, 
one can show that the following FRW form of the metric is
obtained in four dimensions
\be
d s_{4}^{induced}=-d\tau^{2}+r(\tau)^{2} d\Omega_{3,k}^{2}\,.
\ee
The previous form for the projected  metric suggests
an interesting geometrical picture for the evolution
of our four dimensional universe. Since the scale factor in four 
dimensions $r(\tau)$ corresponds precisely to the position
of the brane in the bulk, the motion of the brane along the bulk 
is reflected in the expansion (or contraction) of the corresponding
four dimensional geometry.

Let us consider a brane that contains tension $T_{0}$.
Without entering into details, it is known that in order 
to have a static
brane, that is $r(\tau)=r_{0}$, the junction conditions at the position
of the brane   require, in suitable units, that
\be\label{condten}
T_{0}^{2}-\frac{h(r_{0})}{r_{0}^{2}}=0 \hskip 1 cm\,,\,\hskip 1 cm\,
T_{0}^{2}-\frac{h'(r_{0})}{2\,r_{0}}=0 \,\, .
\ee

\noindent
Once  these conditions are fulfilled, one obtains a
static  four dimensional
model that presents the usual four dimensional gravity.

Now, the idea of \cite{csaki} (for other examples
see \cite{grojean}) is essentially the following:
if the bulk metric presents integration constants  contained
in  its metric coefficient $h(r)$, these integration constants
can be used to compensate, {\it without fine tuning}, the tension
of the brane in formula  (\ref{condten}).

It has been  proved in \cite{clinenogo} that, imposing that the singularity is 
covered by an event horizon, and with   natural requirements
on the bulk field content, this idea cannot be realized in the case of
flat ($k=0$) or negatively ($k=-1$) curved three dimensional subspace. 
This means that
we cannot obtain a static flat solution in four dimensions, but
(possibly) only 
a static, {\it positively curved} solution that does not depend on the value 
of the brane tension $T_{0}$.
In the case  $k=1$, indeed,
the self  tuning mechanism
 works already with the simple Schwarzschild black hole, and we
will limit our discussion on this case.
Taking
\be
h(r)=1-\frac{2M}{r^{2}}
\ee
one fulfills the requirements (\ref{condten}) if
\be\label{condi2}
T_{0}^{2}=\frac{1}{2 r_{0}^{2}} \hskip 1 cm \,{\rm and}\, \hskip 1 cm 
\frac{4M}{r_{0}^{2}}=1\,\, .
\ee
Changing the brane tension, the mass $M$ of the black hole (which
is an integration constant) and the parameter $r_{0}$
change in such a way that the  expressions in (\ref{condi2})  are satisfied; 
at the same time, the volume of the extra dimensions varies its size.
 Consequently, it is easy to see (see for example  \cite{csaki}) that
in this model the effective four dimensional Planck mass
depends on the brane tension {\it via} the following
expression
\be\label{varasymm}
M_{Pl}^{2}(T_0) =\,c\, \frac{M_{f}^{6}}{T_{0}} \,\, ,
\ee
where
$c$ is a(n unimportant) numerical factor, and
 $M_{f}$ is the fundamental scale in five dimensions. Also in this model,
as in the previous ones, a self tuning mechanism is
obtained at the price of changing the Planck mass at the four dimensional
level whenever the brane tension changes.

\section{Self tuning 
in six dimensions and four dimensional physics}\label{consequences}

After a survey of various examples that provide higher dimensional
mechanisms to cancel the cosmological constant, and after commenting
on their effects at the level of four dimensional 
effective physics, we dedicate this section
to a more detailed analysis of the six dimensional model
 we described in Subsection (\ref{asixd}).

This six dimensional approach has various features that render it particularly
 interesting. First, it is possible 
to obtain, without invoking unconventional physics, four dimensional
effective models that present flat three dimensional subspace ($k=0$), 
a geometry that is favored, for 
example, by the recent WMAP observations \cite{wmap}.
Second, it is particularly suitable for interesting generalizations
that allow embedding in supergravity frameworks, as in 
\cite{cliff,gibbons,new}.

In the first subsection, we will explicitly discuss
how four dimensional physics  depends on the brane tension,
 showing that, by a suitable Weyl rescaling,  
it is possible to transfer the dependence on
$T_0$ from the Planck mass to other four dimensional 
quantities. In the second subsection, we will present a qualitative
discussion of the possible contributions, from both brane and bulk 
physics, to the brane tension.

The third subsection, finally, is dedicated to discuss the issue
of the quantization condition that  the form field, present
in this model, must satisfy. We interpret this condition
as a stabilization  mechanism of the extra dimensions,
that, as expected,  results in conflict with the self tuning mechanism.
Integrating out the extra dimensions,
 we present 
a  four dimensional effective description  of the self tuning model,
 based on a non-dynamical 4-form field with a quantized vacuum expectation
value.

\subsection{Variation of constants}\label{variazione}

In this section, we will discuss
in more detail the dependence of four dimensional quantities 
on the brane tension $T_0$. As we will see, 
 a Weyl rescaling  of the metric allows to transfer the $T_0$
dependence 
of the Planck mass to other four dimensional quantities  
potentially measurable.

We know that the six dimensional action (\ref{sixdtotac}) contains
a bulk part that describes gravity, a 2-form field, and a bulk cosmological
constant.
In addition, there is a localized, brane part that is generally modelled
as pure tension. In this section, we are interested in the four dimensional
effective action, obtained after integrating out the 
heavy Kaluza-Klein modes, and keeping only the fields that couple to 
physics localized on the brane. Imposing a suitable
compensation between higher dimensional quantities (see (\ref{dkappa})),
that ensure a {\it flat} four dimensional background, we can model
the four dimensional effective action in the following way

\be\label{fourdstring}
S_{4d}=M_{Pl}^{2}\,(T_0)
\int\, d^{4}x \,\sqrt{-g}\,  \mathcal{R}_{4}  \,
+\int\, d^{4}x \,\sqrt{-g}\,{\cal L}_4
\ee

\noindent
where $M_{Pl}(T_0)$ is the effective four dimensional Planck mass, and
 we introduced a Lagrangian  ${\cal L}_4 \equiv {\cal L}_4(\f,
g_{\mu \nu})$ . This  describes fields
localized on the brane, that we generally denote with 
$\phi$,  that  contribute, {\it via} their dynamics, to 
the brane tension $T_0$. The action (\ref{fourdstring}) explicitly depends
on the brane tension through the four dimensional Planck mass, as

\bea\label{pleff}
M_{Pl}^2\,(T_0) 
 =\left[1-{T_0 \over 4 \pi M_f^4}\right]\,M_{Pl}^2\,(0)\,.
\eea
[$M_{Pl} (0)$ is the Planck mass in absence of branes]

\smallskip

The previous formula connects, in a well defined way, 
the self tuning process
to an observable phenomenon like the variation of the Planck mass.
However, it is interesting to observe that a Weyl transformation
allows  to transmit all the dependence on $T_0$
from the Planck mass to pure four dimensional quantities contained in
 ${\cal L}_4$. Indeed, consider 
the Weyl rescaling
of the metric 
 $g_{\m\n}={1 \over
\a}~\tilde{g}_{\m\n}$, with $\alpha$  given by
\be\label{alfa}
\alpha=1-{T_0 \over 4 \pi M_f^4}\,.
\ee

\noindent
Then the action
 becomes
\be\label{ein4dac}
S^{4d}_{tot}=M_{Pl}^{2}\,(0)\int\, d^{4}x 
\,\sqrt{-\tilde{g}}\, \tilde{{\cal R}}_{4} \,+\int\, d^{4}x 
\,\sqrt{-\tilde{g}}\,\tilde{{\cal L}}_4(\a)\,\,.
\ee
It is clear that all the dependence on $\alpha$ (and consequently on 
$T_0$), has been transmitted 
to $\tilde{{\cal L}}_4$, that is, to the fields localized on the brane.

Interestingly,  whenever $\tilde{{\cal L}}_4$ results
independent on 
$\alpha$, the four dimensional action does not depend on the brane
tension. But this happens only when the original
 ${\cal L}_4$ is scale 
invariant, in which case the cosmological constant problem does not
appear at all.

Choosing a particular ${\cal L}_4$, for the sake
of definiteness,
we can understand more explicitly what
is going on. Consider a simple model in which a gauge symmetry is
spontaneously broken by the $vev$ of some scalar.  A possible
Lagrangian, that includes also fermionic fields, is the following
\be\label{modifilag4}
{\cal L}_4=-{1 \over 4 g_e^2}F_{\m\n}^2 
- \bar{\psi}i\g^{\m}(\de_{\m}-iA_{\m})\psi-|(\de_{\m}-iA_{\m})\f|^2 -
 \xi_0(\f^2-v_0^2)^2-\l_0\f\bar{\psi}\psi\,.
\ee
where $\f$, $\psi$ and $A_{\m}$ are a Higgs scalar, a spinor
 and a gauge boson respectively,   $g_e$
 is the gauge coupling,  $\xi_0$ and $v_0$  the 
Higgs quartic coupling and \vev, and $\l_0$ the Yukawa  coupling.

Performing the abovementioned conformal transformation,
the four dimensional brane action, after rescaling the  fields
 $\psi=\a^{3/4}~\tilde{\psi}$ and $\f=\sqrt{\a}~\tilde{\f}$ 
in such a way that they have canonical kinetic terms, becomes
\be\label{tildeL}
\tilde{{\cal L}}_4=-{1 \over 4 g_e^2}F_{\m\n}^2 
- \bar{\tilde{\psi}}i\g^{\m}(\de_{\m}-iA_{\m})\tilde{\psi}-|(\de_{\m}
-iA_{\m})\tilde{\f}|^2 - \xi\left(\tilde{\f}^2-v^2(\a)\right)^2
-\l \tilde{\f}\bar{\tilde{\psi}}\tilde{\psi}\,,
\ee
where the new constants appearing in the Lagrangian are:
\be
\xi=\xi_0~~~,~~~ \l
=\l_{0}~~~,~~~v(\a)=\frac{v_{0}}{\sqrt{\a}}\,\,.
\ee

Thus, after the Weyl rescaling,
various four dimensional  ``constant'' quantities  acquire a dependence
on $\alpha$, and, consequently, on $T_{0}$.  The masses $m$ of the scalar,
  the spinor  and the vector  fields depend on $\a$ as:
\be\label{massal}
m=\frac{m_0}{\sqrt{\a}}
\ee

\noindent
where $m_0$ denotes the value obtained with
 $T_{0}=0$.
For example,  let us consider how one of the  the masses  varies with the brane
tension. We obtain the formula

\be\label{masferm}
\frac{ \delta m_{\psi}}{m_{\psi}}=\frac{\delta T_{0}}{8 \pi M_{f}^{4}}
\,{1 \over 1-\frac{T_{0}}{ 4 \pi M_{f}^{4}}}\,.
\ee

\subsection{Contributions to the vacuum energy}

After the discussion of the previous subsection, that indicates
how four dimensional quantities depend 
on the brane tension $T_0$ in this model, the next natural
 question to ask 
is which kind of contributions
the brane tension can receive. 
When the contributions are too large, one indeed expects visible effects
at the four dimensional level.

We will not consider a   precise model of fields localized
on the brane
that, {\it via} their dynamics,
contribute to the brane tension: we will limit the discussion
to a qualitative level. Nevertheless,
our considerations allow to draw some conclusions on the characteristics
of the models that can be phenomenologically acceptable.

\vskip 0.3 cm
\noindent
{\it Contributions from quantum corrections}
\vskip 0.3 cm

\noindent
A discussion of the contributions due to quantum 
corrections is necessarily model dependent,
in the sense that it relies on the matter content of the brane
and on the symmetries that one imposes  on the system, both
at the brane and at the bulk level.

In general, one expects that one loop  contributions
to the brane tension that involve brane fields, or Kaluza-Klein (KK)
 modes, result of the order of $M_f$. Possible loop 
factors can in principle lower this quantity, 
but usually, the presence  of a large number of fields
brings back the total contributions to larger values;
therefore, one should study in great detail
quantum corrections in each specific model
to calculate the precise amount of contributions to 
the brane tension.
In any case, it is important to notice that 
 when the size of the contributions becomes too large (close to $4\pi M_f^4$),
formula (\ref{pleff}) tells us that we reach a regime with  strongly
coupled gravitational field, that  is not possible 
to manage.

\smallskip

Imposing additional symmetries
to the system, like {\it e.g.}
 supersymmetry, it is conceivable that quantum  contributions
to $T_0$ can be lowered to a small scale. For example, consider
a supersymmetric model on
the brane, in which supersymmetry is broken
to a scale
$M_{susy} \ll M_f$. This condition
ensures that loop contributions,
involving fields localized on the brane,
 maintain the scale of $T_0$ much below
the  scale $M_f$, at the price, however, of 
introducing new scales in the model. However, loop contributions
involving Kaluza-Klein (KK) modes of bulk fields
 are still potentially dangerous, 
since their expected corrections are still of order $M_f$, the fundamental 
scale in the bulk. 
These contributions
can in principle be limited if one manages to maintain some
powerful symmetry in the bulk, like
supersymmetry and/or some form of scale 
invariance~\cite{cliff,new}.  In this situation, indeed,
the higher
dimensional symmetry reduces 
 the contributions to the brane
 tension due to loops involving KK fields.

We see in this way that, at least conceptually,  it is  possible
to maintain the contributions to $T_0$ below dangerous values. The
 price to pay is the requirement to understand 
how the delicate interplay between
brane and bulk symmetries  can effectively be realized: 
further analysis on specific models are surely required for a complete
comprehension of this point.

\vskip 0.3 cm
\noindent
{\it Contributions from phase transitions}
\vskip 0.3 cm

\noindent
Other possible contributions  to the vacuum energy
come from phase transitions that occur during
the cosmological evolution
of the Universe.
 It is a well known fact  that a phase
transition transmits to the vacuum energy a contribution
proportional to the scale in which it occurs. Although the explicit 
study of cosmology
in this class of six dimensional
models is still in its infancy~\cite{cline6d},
we can nevertheless draw some qualitative observations.

If the phase transition happens at a scale comparable
with the fundamental scale of the theory, $M_f$,
one expects a drastic  change in the Planck mass
(see formula
(\ref{pleff})). 
If, instead, one constructs models in which the typical scale 
for
phase transitions is much lower $M_f$, the change in the Planck mass
is safely small.

\subsection{Quantization conditions, stabilization,
 and four dimensional effective model}

In the previous section we have outlined how
it is possible to conceive 
self tuning models that, although requiring a change in the
volume of the extra space, can limit the effects to observable
physics to an acceptable amount.
In this section, we consider  an additional aspect
of the six dimensional model we are examining, related 
to the issue of stabilization of the extra dimensions.

\smallskip
In \cite{navarro2} (see also
\cite{cliff}), it has been noticed that the charge of the monopole,
relative to the 2-form,
must satisfy a quantization condition,  that ruins the properties
of the bulk action which are used to implement  the idea 
of self tuning~\footnote{The conflict between quantization conditions
and self tuning models have been already discussed, in a different context, in
\cite{dealwis}.}.
This
 fact  {\it per se} is  not surprising:
indeed, it is well known, and largely used in string
theory, that fluxes of form fields through compact spaces
can provide
efficient stabilization
mechanisms of the extra dimensional space. 
Consequently, the six dimensional model presents in its own definition
a   
{\it natural stabilization
mechanism} of the volume, provided by the quantization condition,
that
{\it contrasts} with the requirement  of the self tuning model.

\smallskip

Similar observations have been already pointed out, in a different 
context, in the discussion of four dimensional
models, that rely 
on the introduction 
of a non-dynamical 4-form field, whose value cancels pure cosmological constant
contributions \cite{abbott,BT}.  
Let us briefly summarize the model and these observations,
since they turn out quite useful to understand
the (dimensionally reduced) six dimensional model.

\vskip 0.3 cm
\noindent
{\it The 4-form mechanism in four dimensions}
\vskip 0.3 cm

\noindent
The action for the four dimensional system we consider is
 
\be\label{fourformfourd}
S=\int d^{4}x \, \sqrt{-g}\left[ M_{Pl}^{2} R_4-\frac{1}{48} F_{4}^{2}
-\L_{bare} \right]\,+ {1 \over 6}\int d^{4}x\,\de_{\m}(\sqrt{-g}F^{\m \n \r \s}A_{\n \r \s})  \,, 
\ee

\smallskip 

\noindent
where $\Lambda_{bare}$ represents the various
  contributions to the cosmological
constant. In the above action 
we have added a boundary term, necessary to render
the action  stationary under variations 
that leave $F_4$ fixed on the boundary.
 The solution of the equation of motion for the 4-form reads
\be
F_{\mu \nu \rho \sigma}=c \, \epsilon_{\mu \nu \rho \sigma}\,,
\ee
where $c$ is a constant, and $\epsilon_{\mu \nu \rho \sigma}$ is the four 
dimensional volume density.
 Since $F_{4}$ is a non dynamical field, inserting back this value
in the action one realizes that it is possible to compensate
the value of $\Lambda_{bare}$ via a careful choice of $c$, in such
a way that the effective cosmological constant results as
\be\label{eff4form}
\lambda_{eff}=\Lambda_{bare}+\frac{c^{2}}{2}\,.
\ee

\bigskip

 However,
it has been observed that, embedding the model
in a higher dimensional theory like string/M-theory, the value
of the four form field
is quantized~\cite{BP}, and that the quantization condition
is generally  related to the stabilization of the 
moduli~\cite{garriga,banks,donogue}.
One finds in particular that 
$c_n=e_4 \,n$, where $e_4$ is the 
gauge coupling of the four-form in four dimensions, 
and $n$ an integer number.  Consequently,  only the
part of brane cosmological constant that satisfies 
the following quantization condition can be compensated:

\be\label{leffzero}
\Lambda_{bare}=-n^2\,{ e_4^2 \over 2}\,.
\ee

 When
the steps between  two quantized values of $c$
are sufficiently small, one can in principle obtain a degree
of cancellation that brings $\lambda_{eff}$  inside the interval
fixed by experimental bounds. 
This goal, however, is quite difficult to 
obtain, and in general requires some form of anthropic considerations
\cite{BP,garriga,banks,donogue} (see, however, also \cite{FMSW}).

\vskip 0.3 cm
\noindent
{\it A four dimensional  description of the self-tuning model}
\vskip 0.3 cm

\noindent
The connection  between the 4-form approach in four dimensions
and the 2-form  model in six dimensions, that we have just outlined,
 represents  more than a simple analogy. 
Let us, indeed,  recall the six dimensional action  for the model
of Section (\ref{asixd})
\be\label{azione2form}
S_{tot}=\int d^6x\, \sqrt{-g_6}\left[M_f^4{\cal R}_6-\L_B 
-{1\over 4}F_{2}^2\right]-T_0\int  d^{4}x\sqrt{-g}\,+\int d^{4}x \sqrt{-g}\,{\cal L}_4\,,
\ee
where we  additionally add 
 matter Lagrangian on the brane (consider, as an  example, 
the Lagrangian of \Ref{modifilag4}), while we understand
from now on possible surface terms like the ones present in
(\ref{fourformfourd}). 
 Let us write the previous action in terms 
of the {\it dual} of the 2-form, that in six dimensions
corresponds to a 4-form, $F_4$:
\be\label{azione4form}
S_{tot}=\int d^6x \sqrt{-g_6}\left[M_f^4{\cal R}_6-\L_B 
-{1\over 48}F_{4}^2\right]
-T_0\int  d^{4}x\sqrt{-g}+\int d^{4}x \sqrt{-g}\,{\cal L}_4\, ,
\ee

\noindent
The two systems described by (\ref{azione2form}) and (\ref{azione4form}) are
 equivalent~\cite{teitelboim,BP}~\footnote{See also 
\cite{HML} for a discussion of this issue in  this particular 
system. We thank H.~M.~Lee and S.~F\"orste for valuable 
conversations regarding this point.}.
 We find more comfortable to work, in this context,
with the 4-form field $F_4$.

\smallskip

 At this point, 
let us dimensional reduce this system  from six to four dimensions
integrating out the extra dimension.
 According to \cite{cline6d} the only relevant degrees of freedom 
at low energies are the ones of the massless four dimensional
 graviton (notice that there is no massless field
associated to the deficit angle). Taking into
 account also the  the 
non-dynamical 4-form field,
 one obtains the following dimensionally reduced action:
\be\label{project}
S_{4}=\int d^4x \sqrt{-g} \alpha\left[M^{2}_{Pl}(0) R_{4}
-{1\over 48}\hat{F}_{4}^2-\L  \right]\,+\int d^{4}x \sqrt{-g}\,{\cal L}_4\,,
\ee

\noindent
where $\L=4 \pi R_{0}^{2} \left(\L_B-{2 M_f^4 \over R_0^2}\right)$,
and we define $ \hat{F}_{4} \equiv 2 \sqrt{\pi} R_{0} F_{4}$.
 Notice that the previous effective
 action depends on $\alpha$ {\it via} the explicit
overall factor, and it contains the non-dynamical field $\hat{F}_4$.
The Weyl rescaling we have presented in section
(\ref{variazione}), that is $g_{\mu \nu}=\frac{1}{\alpha} \tilde{g}_{\mu \nu}$,
allows to rewrite the action (\ref{project}) in the following form,
that removes the $\alpha$ dependence from the effective Planck mass:

\be\label{projectsec}
S_{4}=\int d^4x \sqrt{-\tilde{g}_4} \left[M^{2}_{Pl}(0) \tilde{R}_{4}
-{1\over 48}\tilde{F}_{4}^2
-\tilde{\L}(T_0)  \right]\,
+\int d^{4}x \sqrt{-\tilde{g}}\,\tilde{{\cal L}}_4(\alpha)\,.
\ee

\smallskip
\noindent
In the previous formula,
 we define $\tilde{F_{4}}=\alpha^{3/2}\,\hat{F}_{4}$
 to obtain a canonically normalized field
kinetic term, and in addition
$\tilde{\Lambda}(T_0)=\frac{\L}{\alpha}={4 \pi R_{0}^{2}
 \over \a}\left(\L_B-{2 M_f^4 \over R_0^2}\right)$. The 
Lagrangian $\tilde{{\cal L}}_4$ is the one
 given in \Ref{tildeL}, and the parameters appearing in it  
depend on the brane tension.

At this point, comparing the first  
part of (\ref{projectsec}) with the action in (\ref{fourformfourd}),
 it is easy to notice the relation between
the lower dimensional description of the six dimensional mechanism,
and the classical 4-form model. 
Also in this case, keeping fixed the parameter $\Lambda_{B}$,
a change on $T_0$ must be compensated by a change in 
the integration constant relative to the non-dynamical
$F_4$ field. However,
this  form must obey 
  a  quantization condition
that ultimately  corresponds a stabilization
 condition for the volume of the two  extra dimensions.
For our 
 4-form background $\hat{F}_{\mu \nu \rho \sigma}=E \, \epsilon_{\mu \nu \rho \sigma}$  (and accordingly $\tilde{F}_{\mu \nu \rho \sigma}={E
\over \sqrt{\a}}  \, \tilde{\epsilon}_{\mu \nu \rho \sigma}$), 
the quantization condition can be  read from the quantization 
 of the dual 2-form, expressed in \cite{navarro2}. It is given by 
\be
E_n={\sqrt{\pi}  e_6 \over R_0 \a}n
\ee
where $e_6$ is the gauge coupling of the 4-form in 
six dimensions (and it is related to the gauge coupling of the
 4-form in four dimensions by $e_4={ \sqrt{\pi} e_6 \over R_0 \a^{3/2}}$). 

 The effective cosmological constant then has a different 
dependence on the bare cosmological constant (here tension of the brane) than the one noted in (\ref{eff4form}). In this case we obtain
\be
\lambda_{eff}=\frac{1}{\a}\left(\L+{E_n^2 \over 2}\right)\,.
\ee
where the tension dependence (through $\a=1-{T_0 \over 4 \pi M_f^4}$)
 enters both as an overall factor and 
 in the  value of $E_n$.
It would be interesting to construct cosmological scenarios
based on this quantized model, on the lines of the cosmology of the
4-form field \cite{BP,garriga,donogue}.

\section{Conclusions}

We have explored the self tuning idea in brane world models based on extra
spacetime dimensions. These are models where solutions with a Poincar\'e
invariant vacuum can be found irrespectively of the brane  vacuum energy
(tension), when keeping all the other bulk and brane parameters fixed.  We have analyzed the various ways that self tuning can be
realized and reviewed explicit examples where this can be achieved.

The important message of the present paper is that, in general, the  four
dimensional physics {\it is affected} by the modification of the brane
vacuum energy. Indeed, all the avaliable examples present a variation of
the compactification volume as the tension of the brane changes. This
means that, although solutions can be found  where the curvature of
the four dimensional space remains zero as the  brane tension changes, the effective Planck mass or the masses  of 
matter fields on the brane  {\it vary}. Consequently, it is important
to determine, in discussing a specific model, 
 whether  the variation of observable quantities can be limited 
in order to satisfy phenomenological bounds.

In addition,  it is  observed           
that a generic feature of the self tuning
mechanism is the variation of the compactification volume as a function of
the brane tension. In the examples we discussed, any mechanism that
would be used to stabilize the size of the extra dimensions will be in 
apparent conflict with self tuning: once the extra dimension is stabilized
the brane vacuum energy cannot vary without re-introducing fine tuning.

The most promising example of self tuning that we have singled out in
 the paper is the one with three-branes embedded in six dimensions.
An interesting property of co-dimension two branes, firstly noticed in
\cite{luty}, makes this model particularly attractive. However, the
quantization condition of the gauge field used in the particular setup,
that  essentially corresponds to 
a  stabilization condition for the extra dimensions, disturbs the self
tuning. For the particular model,  
 we  show that the six dimensional 
 self tuning resembles  a higher dimensional
realization of a well known   mechanism to compensate  the 
cosmological
constant
 in four dimensions, based on a non-dynamical field.

\smallskip
We would like to stress that 
 the self tuning
models have opened a new point of view to the  cosmological constant
problem, but still there are various issues
 that deserve further investigations.
 The variation of  four dimensional constants should be treated with
caution since, as we noted, in some cases one 
 needs to keep the  contributions to the vacuum
energy  smaller than the higher dimensional fundamental scale. This
could imply   the necessity to 
 introduce new scales on the model, or invoking 
additional symmetries. In addition, the apparent conflict between self
tuning and stabilization should be taken into account when  considering,
in a specific model, a mechanism to stabilize the extra dimensions.
 Furthermore, the question of how to eliminate 
 nearby curved
solutions remains  an important challenge.

\smallskip

Concluding, the self tuning brane world models provide a new insight to
the cosmological constant problem and are accompanied with very
interesting four dimensional signatures, as well as potential disturbing
problems. A more careful study of these models, that includes also 
an analysis of the dynamics of matter localized on the brane,
 may shed light on many of the unsettled  aspects of these
models.

\acknowledgments{ We would like to acknowledge helpful discussions with
J.~Cline, S.~F\"orste, D.~Grellscheid,
H.~M.~Lee, I.~Navarro, S.~Parameswaran, F.~Quevedo and I.~Zavala.
We thank S.~F\"orste for pointing out a minor
 error in the first version
of the preprint.
 This work is supported in part by the European Community's Human Potential
Programme under contracts HPRN--CT--2000--00131 Quantum Spacetime,
HPRN--CT--2000--00148 Physics Across the Present Energy Frontier
and HPRN--CT--2000--00152 Supersymmetry and the Early Universe. We dedicate this work to the memory of Ian I. Kogan.
}

\appendix

\def\theequation{A.\arabic{equation}}
\setcounter{equation}{0}
\section{The singular part of the scalar curvature}\label{appSing}

In this appendix we present the relation of the singular part of the curvature scalar to the brane tension in $D$ dimensions, obtained by the Einstein equations. Suppose we start from an action of the form:
\be
S_{tot}=M_{f}^{D-2}\int d^{D}x \sqrt{-g_{D}}\mathcal{R}_{D}+S_{B,br}
\ee
where $S_{B,br}$ is the action of the bulk and the branes present in the model. The Einstein equations derived from the above action are:
\be
{\cal R}^D_{MN}-{1 \over 2} g^D_{MN}{\cal R}^D={1 \over 2 M_f^{D-2}}~T_{MN}
\ee
and their trace is:
\be
 M_f^{D-2}{\cal R}_D=-{1 \over D-2}T_M^M \label{traceE}
\ee

Suppose the $D$-dimensional metric has the form  \Ref{Dmetric}:
\be
d s_{D}^{2}=e^{2 W(y)}~ g_{\mu \nu}(x)\, d x^{\mu} d x^{\nu}
+\g_{m n}
(y)d y^{m} d y^{n}
\ee
where we have in general the $x^{\m}$ coordinates to span a $(p+1)$-dimensional subspace. If the model has only $p$-branes along the $x^i$ directions, then the energy momentum tensor is:
\be
T_{MN}=T_{MN}^{(B)}-\sum_i e^{2 W(y)}~ g_{\mu \nu}(x)e^{\s_i \phi(y)}T_i \D^{(d)}(y-y_i)
\ee
where  $T_{MN}^{(B)}$is the bulk contribution to the energy-momentum tensor and the next terms are the ones of the $p$-branes coupled to a scalar field $\f$. We denote by  $\D^{(d)}(y-y_i)$ the generalized delta function with the property $\int \sqrt{\g}\D^{(d)}(y-y_i)=1$ \cite{myers}. Taking the trace we find:
\be
T_M^M=T_{~~~~M}^{(B)~M}-(p+1)\sum_i e^{\s_i \phi(y)}T_i \D^{(d)}(y-y_i)
\ee
where the factor $(p+1)$ appears because the trace of the induced metric in the $(p+1)$ longitudinal dimensions of the $p$-brane has been performed. Combining the above equality with \Ref{traceE}, we find that  the singular part of the higher dimensional curvature is:
\be
M_f^{D-2}{\cal R}_D^{sing}={p+1 \over D-2}\sum_i e^{\s_i \phi(y)} T_i \D^{(d)}(y-y_i)
\ee
Hence, the quantities $R_i^{sing}$ defined in section \ref{sftuneaction} are:
\be
M_f^{D-2}R_i^{sing}={p+1 \over D-2}e^{\s_i \phi(y_i)} T_i
\ee

\end{document}